\RequirePackage{amsthm,amsmath,amssymb,amsfonts}
\documentclass[sn-mathphys,Numbered,iicol,pdflatex]{sn-jnl}

\usepackage{graphicx}%
\usepackage{multirow}%
\usepackage{amsmath,amssymb,amsfonts}%
\usepackage{amsthm}%
\usepackage{mathrsfs}%
\usepackage[title]{appendix}%
\usepackage{xcolor}%
\usepackage{textcomp}%
\usepackage{manyfoot}%
\usepackage{booktabs}%
\usepackage{algorithm}%
\usepackage{algorithmicx}%
\usepackage{algpseudocode}%
\usepackage{listings}%
\usepackage{adjustbox}
\usepackage{siunitx}
\usepackage{subcaption}
\usepackage{orcidlink}
\usepackage{url}

\usepackage{dcolumn}
\newcolumntype{d}{D{.}{.}{-1}}
\newcolumntype{f}[1]{D{.}{.}{#1}}

\newcommand{\eg}{{e.g., }}
\newcommand{\ie}{{i.e., }}

\theoremstyle{thmstyleone}%
\theoremstyle{thmstyletwo}%

\theoremstyle{thmstylethree}%

\begin{document}

\title[Boron-like sulfur and argon]{High-precision, reference-free measurements of $2p \to 1s$ transitions in boron-like sulfur and argon}


\author*[1,2]{\fnm{Louis} \sur{Duval}\orcidlink{https://orcid.org/0000-0002-1720-1699}}\email{louis.duval@lkb.upmc.fr}
\author[2]{\fnm{Emily} \sur{Lamour}\orcidlink{https://orcid.org/0000-0001-8467-2637}}
\author[2]{\fnm{Stéphane} \sur{Macé}}
\author[3]{\fnm{Jorge} \sur{Machado}\orcidlink{https://orcid.org/0000-0002-0383-4882}}
\author[2]{\fnm{Marleen} \sur{Maxton}}
\author[1]{\fnm{Nancy} \sur{Paul\orcidlink{https://orcid.org/0000-0003-4469-780X}}}
\author[2]{\fnm{Christophe} \sur{Prigent}}
\author[2]{\fnm{Martino} \sur{Trassinelli}\orcidlink{https://orcid.org/0000-0003-4414-1801}}
\author[1]{\fnm{Paul} \sur{Indelicato\orcidlink{https://orcid.org/0000-0003-4668-8958}}}

\affil[1]{\orgdiv{Laboratoire Kastler-Brossel}, \orgname{Sorbonne Université, Collège de France, CNRS, École Normale Supérieure}, \orgaddress{\street{4 place Jussieu}, \city{Paris}, \postcode{75005}, \state{France}.}}
\affil[2]{\orgdiv{Institut des NanoSciences de Paris}, \orgname{Sorbonne Université, CNRS}, \orgaddress{\street{4 place Jussieu}, \city{Paris}, \postcode{75005}, \state{France}.}}
\affil[3]{\orgdiv{Laboratório de Instrumentação Engenharia Biomédica e Física da Radiação}, \orgname{Facultade Cîencas e Tecnologia, Universidade Nova de Lisboa}, \orgaddress{\city{Caparica}, \postcode{2829-516}, \country{Portugal}}}


\abstract{We have measured several $2p \to 1s$ transition energies in core-excited boron-like ions of sulfur and argon. The measurements are reference-free, with an accuracy of a few parts per million. The x-rays were produced by the plasma of a an electron-cyclotron resonance ion source and were measured by a double-crystal x-ray spectrometer. The precision obtained for the measured $1s 2s^2 2p^2~J - 1s^2 2s^2 2p~J'$ lines is \qty{\approx 4}{ppm} for sulfur and \qty{\approx 2}{ppm} for argon. The line energies are compared to relativistic atomic structure calculations performed with the \textsc{mdfgme} multi-configuration Dirac-Fock code. This comparison is used for line identification and test the theoretical methods, which reach an agreement with experimental data up to \qty{15}{\milli\electronvolt}. The theoretical calculations have been extended to C$^+$, Si$^{9+}$, Cr$^{19+}$ and Fe$^{21+}$, which were the only B-like ions where such transitions were measured up to now.    }




\maketitle

\section{Introduction}\label{sec1}
The experimental study of few-electron, highly charged ions (HCI) has consistently improved in accuracy across a broad range of elements (see \eg \cite{ind2019} for a review of one to three electron ions). Measurements of transitions to the $1s$ level in one and two electron ions provide sensitive tests of Bound State Quantum Electrodynamics (BSQED). Transitions between $2p\to 2s$  have also been measured in a number of two and three-electron ions \cite{ind2019} including heavy ions.  In particular, high precision measurements of $2p\to 2s$ transition energies in two to four electron uranium ions have been recently measured \cite{lbds2024}.  At present, the \(1s^2 2s^2 2p~ ^2P_{3/2} -~^2P_{1/2}\) transition wavelength in boron-like argon (five electrons) is the most accurately measured transition in highly-charged ions \cite{mkbl2011,eahk2019}.  Its Landé $g$-factor and upper level lifetime have also been measured \cite{mlkb2020}.  

However, very few measurements of systems with open core shells and three electrons or more have been performed. High-precision, reference-free measurements of $n=2 \to n=1$ transitions in lithium-like and sulfur and argon \cite{mbpt2020} and beryllium-like argon have been published in the last few years \cite{mssa2018}. Theoretical calculations in such systems becomes increasingly difficult with the number of electrons.
In core-excited lines, With a growing number of electrons, more lines can exist, , with nearby energies, and broad Auger widths, increasing the possibility of unresolved transitions. At the same time, calculation of the level energies becomes significantly harder for core-excited ions with more than four electrons. The QED effects are masked by the uncertainties coming from solving the many-body problem.  There are only four elements --- B, C~\cite{kah2022}, Cr~\cite{bfq1999} and Fe~\cite{bpjh1993,rbes2013} --- for which $2p \to 1s$ transitions in the boron electronic sequence are listed in the NIST Database \cite{krrn2023}. The Livermore group also performed measurements of Si and S. \cite{hbwg2016}.

These few electron systems with K-holes are growing in importance as they are now frequently observed by  space-borne x-ray spectrometers, and in particular high-resolution microcalorimeters (see, \eg \cite{lcmm2014}). A new era of high precision spectroscopy of hot plasmas of the universe started with the launch of the soft x-ray microcalorimeter built within Hitomi~\cite{ttkm2016}, back in 2018. However, the failure of the satellite postponed this high precision study of the universe, until XRISM was launched in September 2023. The first observation with this new satellite of the  Supernova remnant N132D has already been published~\cite{xrism2024}. The three most intense elements seen in the spectrum are silicon, sulfur and argon. The precise measurement of K-hole spectra for different ionization states of these elements is useful for online calibration of the satellite. 
Some of these lines have been recently been remeasured by Livermore~\cite{hbwg2016}, using electron-beam ion traps (EBITs) and a microcalorimeter, but both the detector resolution and the ionization method lead to unresolved lines. Crystal spectroscopy offers much better energy resolution, around a few ppm in the \qtyrange{2}{6}{\kilo\electronvolt} energy range, with the drawback of accessing a only narrow window of energy, requiring many measurements. As already demonstrated in \cite{assg2014,mssa2018,mbpt2020,mpsd2023}, by coupling an electron cyclotron ion source (ECRIS), to a Double Crystal Spectrometer (DCS), it is possible to perform reference-free, ppm level accuracy energy measurements.
Here we present measurements, in the region of the \(1s 2s^2 2p^2~^2P_{3/2} - 1s^2 2s^2 2p~^2P_{3/2}\) transitions in B-like S and Ar.
Such transitions are regularly observed in astrophysical~\cite{hhgb2019} and plasma measurements \cite{bqfs2000,mfsp2002,hbwg2016} but are generally unresolved, making a unambiguous identification impossible. The present measurement reveals a complex structure, which we subsequently interpret with the help of high-precision BSQED calculations.
\section{Description of the experiment}
Only the main features of the experiment will be recalled here as the detailed explanations of the experimental approach are provided in Refs.~\cite{assg2014,mssa2018,mbpt2020,mpsd2023}. The highly-charged ions are produced in the plasma of a permanent magnet Electron-Cyclotron resonance ion source (ECRIS), which is injected with a   \qty{14.5}{\giga\hertz} microwaves with a power of approximately \qty{300}{\watt}. The electronic temperature is around \qty{50}{\kilo\electronvolt} \cite{gtas2010},  far above the $1s$ ionization threshold of  many elements including Xe. The plasma is created from the gas of interest (here SF$_6$  and Ar) and a support gas (typically oxygen), which supplies electrons for ionization. A vacuum double crystal spectrometer (DCS) is connected to the source by an evacuated pipe of length around \qty{1.5}{\meter}. The x rays are observed through  a \qty{50}{\micro\meter}-thick Be window to separate the ECRIS vacuum and the DCS vacuum. The spectrometer is under vacuum to avoid absorption of the low-energy x rays emitted by elements like S and Ar \cite{des1967}. The DCS is precisely aligned with the ECRIS axis to aim at the plasma's center position. The plasma, with a diameter of about \qty{30}{\milli\meter}, is observable by the spectrometer through a copper tube (the polarization electrode of the source) with a diameter of \qty{6}{\milli\meter}.

The crystals used for the DCS are pure silicon (111) crystals, which lattice spacing was measured to a \num{1.2e-8} relative accuracy at NIST \cite{assg2014}.
Photons reflected off the crystals are detected using a cooled, large-area avalanche photodiode (LAAPD). Cooling of the detector is achieved using a water-ethanol bath at \qty{-18}{\celsius}, which minimize electronic noise and energy resolution for x-ray photon measurements~\cite{laab2005}. Since the spectrometer operates in vacuum, the cooling system additionally serves to dissipate heat from the angular tables stepping motors and cool the crystals supports. The temperature of each crystal is measured using calibrated thermistors to better than \qty{0.1}{\celsius}. It is stabilized by a proportional-integral-derivative (PID) controller, which  use the measured temperature to control heating resistors.

During the measurement, the first crystal is kept at a fixed position and acts as an energy selector such that a limited range of photon energies reflect towards the second crystal. 
The second crystal is then oriented into one of two position: (1) nondispersive, reflection of order $(n, -n)$ in Allison's notation~\cite{aaw1930}, where the two crystals are parallel and the photons bounce off both crystals independently of their energies, or (2) dispersive, reflection of order $(n, n)$, where both crystals deflect the photons in the same direction.
The second crystal angle is scanned around both positions, in a sequence nondispersive -- dispersive -- nondispersive, use for each day of measurement.

The experimental campaigns were conducted in summer 2017 and summer 2018 and focused on the \(1s 2s^2 2p^2~^2P_{1/2}-1s^2 2s^2 2p~^2P_{1/2}\) and \(1s 2s^2 2p^2~^2P_{3/2}-1s^2 2s^2 2p~^2P_{3/2}\) transitions in B-like sulfur and argon.  
The measurements were performed in a room with limited temperature regulation, during a hot summer, which induced large temperature variations throughout the measurement period. We thus had to setup our crystal temperature regulation quite above the usual \qty{22.5}{\celsius}, which is the temperature at with the lattice spacing of the crystal has been measured. This also led to fluctuations of the temperature of the crystals larger than in previous measurements.

\begin{figure}[ht]
    \centering
    \caption{Spectra of sulfur and argon with identification. A fit was added for eye's comfort. The angle was converted to energy using the formula from Ref.~\cite{assg2014}. The identification and the comparison with theory is discussed in Sec. \ref{sec:disc-comp-theo}}\label{fig:s_ar_spectra}
    
    \includegraphics[width=\linewidth]{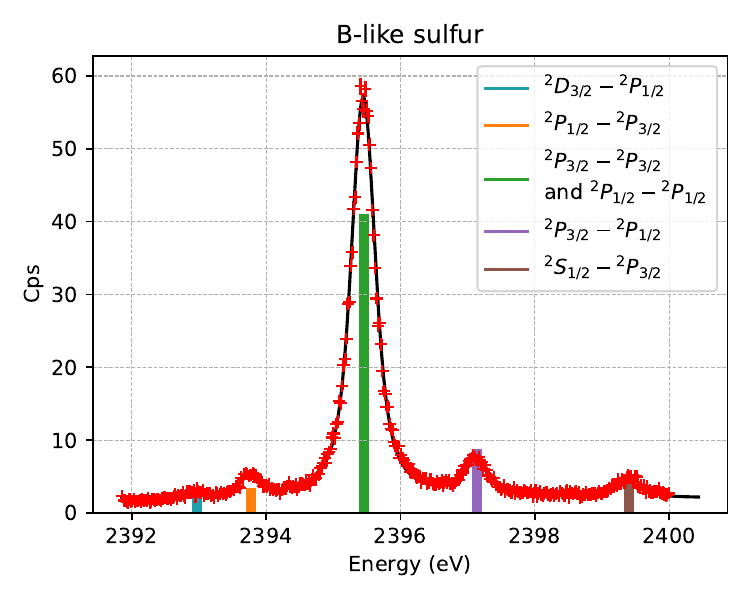}
    \includegraphics[width=\linewidth]{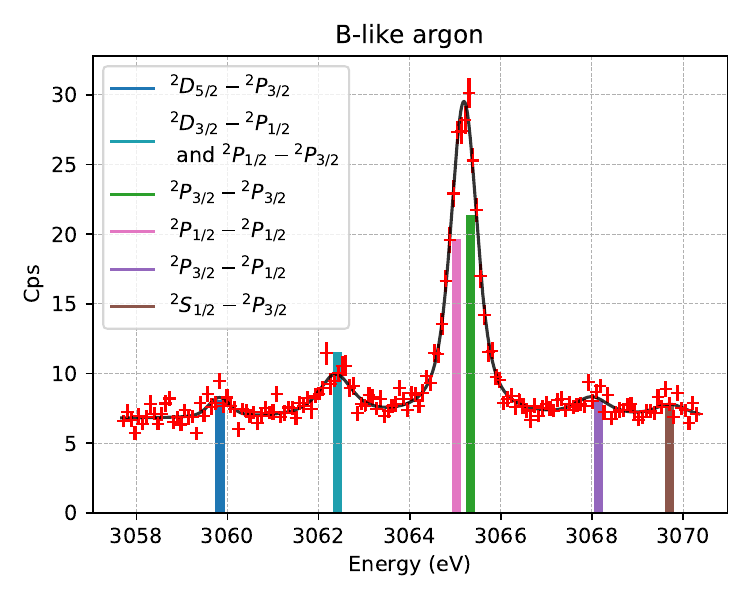}
\end{figure}

\section{Identification of the transitions}
Typical spectra of B-like sulfur and argon are presented in Fig.~\ref{fig:s_ar_spectra}. As one can see, they look quite similar, both composed of five visible peaks. For easier designation, we will refer to each peak as peak one to five, from left to right. At first look, the second and fourth peak seems to be symmetric with respect to the most intense one. In argon, the third peak seems to be composed of two lines, which are not totally resolved by the spectrometer. The argon's second peak has a larger width than the equivalent one in the sulfur spectrum and than the other sulfur peaks.
 
The overall process of the analysis is similar to the one  described in our previous work on He- to Be-like Ar and S \cite{asgl2012,mssa2018,mbpt2020,mpsd2023}, as it uses simulated x-ray profiles, derived from dynamical diffraction theory and exact crystal response function. 
In addition, we use here a Bayesian analysis (BA) method to be able to disentangle the contribution of the different transitions in each spectral peak.  
More precisely we use the Bayesian model selection method \cite{sas2006} to assign probabilities to different data models, and get the associated widths and energies of the measured transitions. Such methods were used in a previous work \cite{mpsd2023} for very specific needs: to check the influence of the different crystal modelization methods and in \cite{mbpt2020} to extract the energy of the M2 forbidden line $1s 2s 2p~^4P_{5/2}-1s^2 2s^2~^2S_{1/2}$ in Li-like argon and sulfur. Differently to our past works, these methods are used here systematically for each step of the data analysis.   

\subsection{Modelling the contributions}\label{subsec:simu}
Before going into the BA and determine how many transitions are present in these spectra, it is crucial to chose properly which kind of model should be used for reproducing the experimental lineshape.
For a first analysis, a toy-model has been used based on Voigt profiles, a convolution product of a Lorentz and a gaussian function. These profiles take into account the Doppler broadening and the broadening due to the reflection curves of the crystals as the gaussian width, and the natural width of the transition as the Lorentz width. In our case, this lineshape is efficient for fitting lower statistic contributions, but cannot describe the most intense peaks as it does not take into account the crystals' response function shape.

Following, in order to take into account all the effects from the setup, profiles based on \textit{ab initio} simulations~\cite{assg2014} are considered to take into account the effects due to the geometry of the spectrometer and the ECRIS\@ and the reflection profile of the crystals. The simulation take as input reflectivity profiles computed with the xraylib~\cite{bsgs2004} and XOPPY~\cite{sad2011} python libraries, for the Si(111) crystals, and physical parameters such as the photons' energy, the Doppler width and the natural width. The Doppler broadening, considered Gaussian, is fixed to the values measured in the M1 transitions of He-like Ar and S, published in Refs.~\cite{mbpt2020,mpsd2023}, which have a negligible natural width with respect to the Doppler width.
More precisely, we used the values \qty{65+-4}{\milli\electronvolt} for argon \cite{asgl2012} and \qty{84+-6}{\milli\electronvolt} for sulfur \cite{mpsd2023}. The energy of the x rays is fixed to \qty{3065}{\electronvolt} for Ar and \qty{2395}{\electronvolt} for sulfur. Different simulated profiles are simulated considering natural widths between \qty{0.01}{\electronvolt} and \qty{0.6}{\electronvolt}, and are compared to the experimental data to deduce the most probable one. 
The width determined through this method is then used for further simulations in order to deduce the energy (see sec. \ref{subsec:energy_determ})  

\subsection{Determination of the number of contributions}\label{subsec:num_contrib}
These spectra (see Fig. \ref{fig:s_ar_spectra}, contains several contributions, with some which may be unresolved. The main challenge is to ascertain properly the number of contributions, in order to determine their widths and energy.  To dissect the contributions within these five peaks spectra, we employed Bayesian model selection methods that, as we will see,  bring several key benefits. 
To infer the number of contributions in each peak, we use a set of models containing different number of contributions. For all the models, we keep the same background modeling and lineshape for each contribution:
\begin{align}
    \label{eq:generic_Model}
    &M\left(\theta,\theta_i,I_i,\sigma,\gamma_i,a\right)=\nonumber\\&\sum_{i=1}^{c_{\max}}I_i*P({(\theta-\theta_i)},\sigma,\gamma_i) + a,
\end{align}
where \(P\) is the lineshape  \(I_i\) is the intensity of the transition, \(\theta_i\) the position of the transition, \(\gamma_i\) the natural linewidth of the transition, \(\sigma \) the broadening due to Doppler effect and instrumental response, and \(a\) the background. Depending on the statistics in the peak, the lineshape used is either a Voigt profile or a simulation. The Voigt profile is used used for the peaks with low statistics (peaks 1, 2, 4, and 5). For peaks with significant statistic such as the third peak of each spectra, for both elements, \textit{ab initio} simulation is required to descibe sufficently well the data.

\begin{figure*}
    \centering
    \caption{Difference of the log Evidence for every peak of argon and sulfur. Every peak seems to have only one atomic contribution in sulfur. For argon, only the third peak exhibit two contributions.  The evidence differences for the third peak is bigger than the others due to larger statistics. }\label{fig:ev_ppp}
    \includegraphics[width=.49\linewidth]{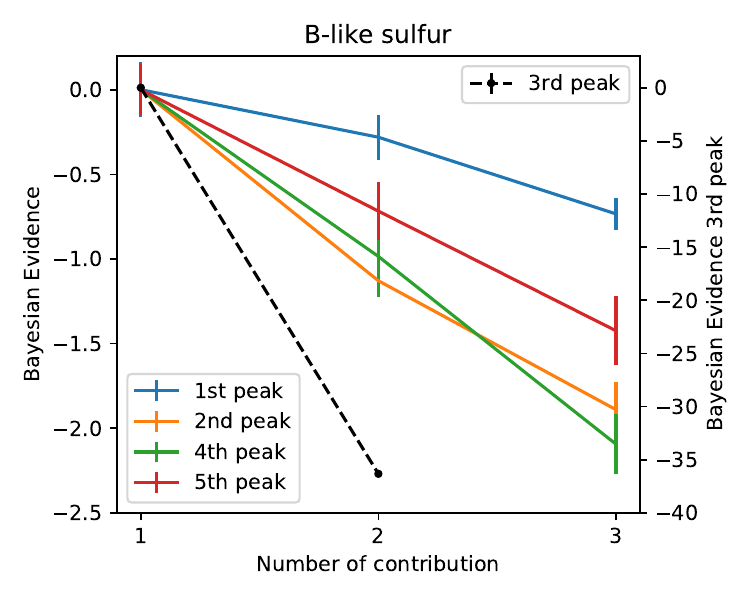}
    \includegraphics[width=.49\linewidth]{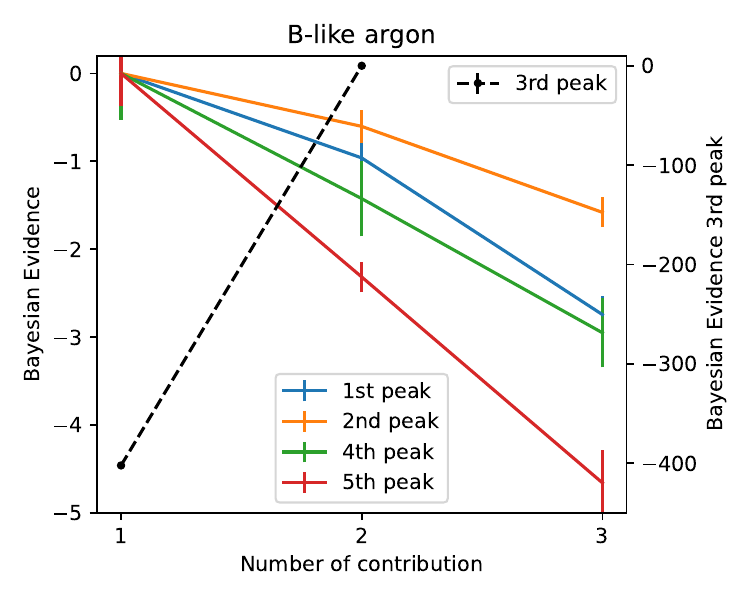}
\end{figure*}

For each simulation, the experimental data are fitted using the code \textsc{nested\_fit} \cite{tra2017,tra2019,mft2023} to deduce the most probable parameters, their uncertainties, and the Bayesian evidence (BE), the marginalized value of the likelihood over the possible parameter values.
From the Bayesian evidence, the probability of each model, \ie the number of unresolved contribution and their linewidths, is deduced.
(More details on the Bayesian model selection can be found in Refs.~\cite{sas2004,tro2008,tou2011}.)
Subsequently, we identify the model with the highest BE for each specified number of contributions, comparing these across different sets defined by their natural linewidths. This method is not applied for the whole spectrum, but to each peak, defining by eye a cutting in between each contribution. The neighboring contributions of the studied peak is modelled with a Voigt profile, to take into account the long tails.

\begin{figure}
    \caption{Example of the second argon peak, fitted with two voigt profiles and a constant background. The green contribution is the tail of the neighboring third peak. }\label{fig:p2_2P}
    \includegraphics[width=\linewidth]{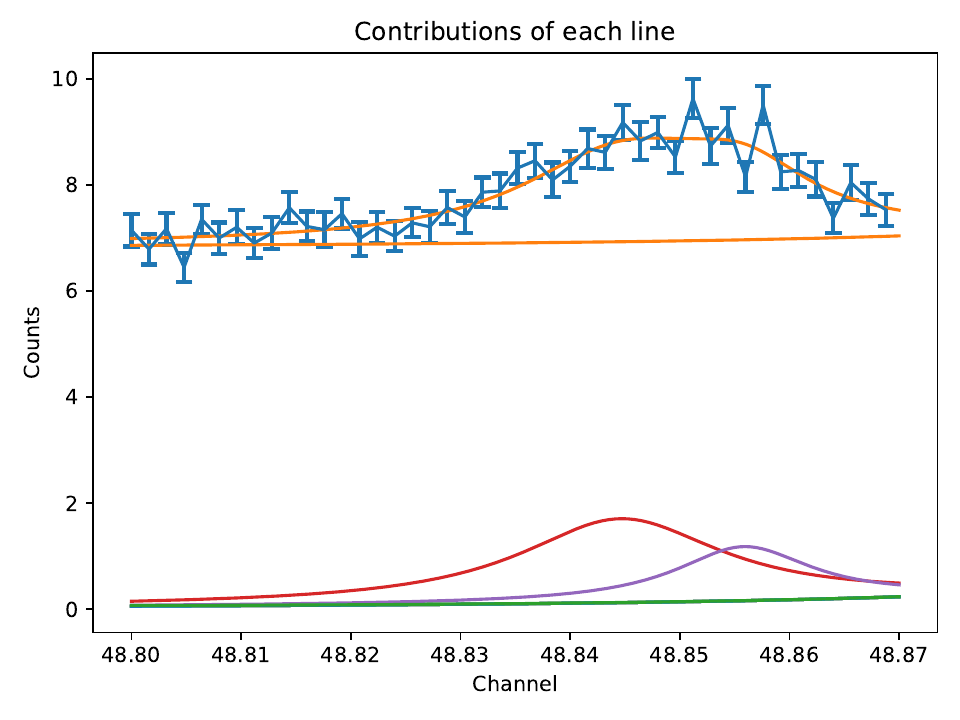}
\end{figure}

For the sulfur spectra, all the peaks show the highest evidence when using a single line simulated profile, indicating, with the present statistics, that there are no line blends and that each peak visible in the spectrum correspond to a single transition (see Fig.\ref{fig:ev_ppp}.  
The same result was obtained for all peaks of the argon spectra except the third one (see Fig.\ref{fig:ev_ppp}.
For this latter one, by using the simulations, the evidence was highest for a two transitions model.
The case of the second peak in argon was peculiar, as it displayed a much larger natural width (cf Fig.~\ref{fig:p2_2P}) than the other peaks with only one contribution. By limiting the natural width below \qty{400}{\milli\electronvolt}, we obtained a BE in favor of a two contribution model, as the one contribution model had the highest evidence for a natural width \qty{\approx 650}{\milli\electronvolt}, significantly different from \qty{\approx 300}{\milli\electronvolt}, the width of all other peaks with one component only.

\subsection{Identification of the transitions}
\begin{figure}[htbp]
    \caption{Transition scheme for the principal peak and its two neighbors. The symmetry of both neighbors can be explained by the fact that two upper level decays into two lower level. As the splitting of both the upper levels and lower levels is around the same, i.e. \(E_1-E_3 \approx E_2-E_4\), the central peak of the spectrum is blended. With \(E_2-E1 = E_4-E_3\), \(E_2\) and \(E_3\) transitions are symmetric with respect to the \(E_1\) and \(E_4\) transitions }\label{fig:level_scheme}
    \includegraphics[width=\linewidth]{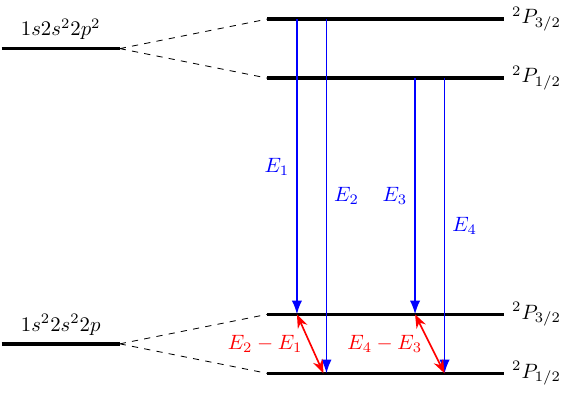}
\end{figure}

The dominant process in the ion source involves K-shell ionization of a ground state ion, which correspond in our specific case to the removal of a K-shell electron from a \(1s^2 2s^2 2p~^3P_0\) carbon-like ion.  We then expect that the main population of K-shell excited B-like ions will be in \(1s 2s^2 2p^2~^2P_{i}\), \(i \in [1/2, 3/2]\). These two levels can decay into two lower sublevels. The most intense transitions will be the two  \(1s 2s^2 2p^2~^2P_{i}-1s^2 2s^2 2p~^2P_{i}, i \in [1/2,3/2]\). The transitions two transitions with $^2P_{i}-^2P_{j},~i\neq j$ are less probable as they require a spin flip of the decaying electron. 
Moreover, the $^2P_{1/2}-^2P_{3/2}$ energy is almost the same for both $1s 2s^2 2p^2$ ans $1s^2 2s^2 2p$ electronic states. Hence the two \(1s 2s^2 2p^2~^2P_{i}-1s^2 2s^2 2p~^2P_{i}, i \in [1/2,3/2]\) transitions will be around the same energy, surrounded by the two \(1s 2s^2 2p^2~^2P_{i}-1s^2 2s^2 2p~^2P_{j}, (i,j) \in [1/2,3/2],~i\neq j\) transitions, which energy differs from the \(1s 2s^2 2p^2~^2P_{i}-1s^2 2s^2 2p~^2P_{i}, i \in [1/2,3/2]\) by the splitting of the $1s^2 2s^2 2p ^2P_{1/2} - $
The transition schemes are shown in Fig.~\ref{fig:level_scheme}. This particularity results in a intense peak, containing two transitions, which are surrounded with two weak lines, symmetric with respect to the intense line.

Moreover, in the argon spectra, the energy difference between the main peak and its adjacent peaks is close to the energy deduced by laser spectroscopy for the \(1s^2 2s^2 2p~^2P_{1/2}-^2P_{3/2}\) transition, measured at \qty{\approx2.81}{\electronvolt}~\cite{shca2006}. The same is observed for sulfur, with the \(1s^2 2s^2 2p~^2P_{1/2}-^2P_{3/2}\) transition, measured at \qty{1.62857}{\electronvolt} by grating spectroscopy of sun emission~\cite{jeff1969}.

As written above, the argon's second peak shows a larger linewidth than the other peaks, indicating the possible presence of an additional transition within this peak. This additional transition is identified as \(1s 2s^2 2p^2~^2D_{3/2}-~^2\)P\(_{1/2}\), which accounts for the existence of the initial peak along with \(1s 2s^2 2p^2~^2D_{5/2} -~^2P_{3/2}\). In sulfur, the \(^2\)D level decreases in energy, pushing the \(^2\)D\(_{5/2}\) beyond the measurement range and separating the \(1s 2s^2 2p^2~^2D_{3/2}-1s^2 2s^2 2p~^2P_{1/2}\) from the \(1s 2s^2 2p^2~^2P_{1/2}-1s^2 2s^2 2p~^2P_{3/2}\). 
Finally, the fifth peak observed in the spectra is identified to be \(1s 2s^2 2p^2~^2S_{1/2}-1s^2 2s^2 2p~^2P_{3/2}\) for both argon and sulfur. The identifications of the peaks for both argon and sulfur are displayed in Fig.~\ref{fig:s_ar_spectra}.

\section{Extraction of the physical parameters}
\subsection{Determination of the widths}\label{subsec:width_determination}
\begin{figure}
    \caption{Fit of the evidence depending on the width of the natural width simulations with a 4-th order polynomial for \(1s 2s^2 2p^2~^2P_{3/2}-1s^2 2s^2 2p~^2P_{3/2}\)}\label{fig:evidence_fit}
    \includegraphics[width=\linewidth]{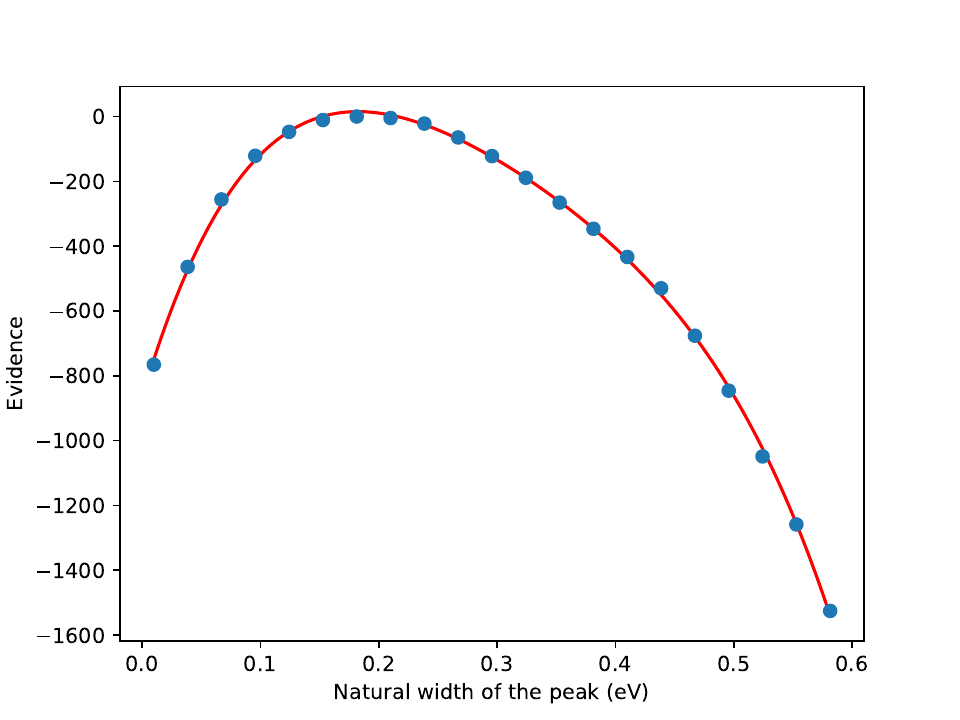}
\end{figure}

The determination of the width is a key step of the analysis, as we require the correct natural width to input in the simulation to extract the final transition energy values (\textit{cf} Sec.~\ref{subsec:energy_determ}).
As the width is not a free fit parameter because of the complex simulation process, we have to run different analyses, one each profile simulated with a different width value. Following, we apply once more the approach of maximum evidence to determine the most probable value of the width. During this procedure, all the spectra are taken into account at once considering the same simulation for all available by considering the total likelihood equal to the product of the single spectra likelihood. In this way the statistical uncertainty of the width determination is drastically. 
For the peaks with one component only, the most probable width value and the associated uncertainty are determined finding the maximum of a 4th degree polynomial used as interpolation of the BE dependency as function of the width. An example of this fit is given in Fig~\ref{fig:evidence_fit}. The maximum gives the measured natural width of the line, while the maximum BE value (in log) minus 0.96, which correspond to \(2\sigma \) uncertainty~\cite{gat2007}. For the peaks with two contributions (in argon only), the evidence is a function of two different widths. In order to get the width of the two contributions, we project the  2D evidence function on the width of the other contribution, by taking the maximum of the evidence: \({BE(w_i)=\max_j(BE(w_i,w_j))}\). We were unable to extract any width for the unresolved \(1s 2s^2 2p^2~^{2}P_{1/2} - 1s2~2s^2~2p~^2P_{1/2}\) and \(1s 2s^2 2p^2~^{2}P_{3/2} - 1s2~2s^2~2p~^2P_{3/2}\) in sulfur and \(1s 2s^2 2p^2~^{2}P_{1/2} - 1s2~2s^2~2p~^2P_{3/2}\) and \(1s 2s^2 2p^2~^{2}D_{3/2} - 1s2~2s^2~2p~^2P_{1/2}\) in argon. All the widths are shown in Table~\ref{tab:widths} for argon and sulfur.

From this analysis, the energy of the different transitions is also deduced by comparing the angle difference between the simulations and each measured transitions. This allowed in addition to crosscheck any unattended correlation between the natural width and an approximation of the transition energy. This energy will be used in the next step for the final energy search.

\begin{table}[ht]
    \centering
    \caption{Measured widths for sulfur and argon. All the transitions are from  the $1s~2s^2~2p^2\, ^{2S+1}L_J$ electronic state to $1s^2~2s^2~2p\, ^2P_J$ state. If two transitions are mentioned, the width shown is the combined width. Values are in \unit{\milli\electronvolt}. Theoretical values are described in Sec \ref{sec:theo-calc}}\label{tab:widths}.
    \begin{tabular}{lll}\toprule 
    Transition	 &	\multicolumn{1}{l}{Exp. } 	& \multicolumn{1}{l}{Th.}\\
    \midrule
    \multicolumn{2}{c}{Sulfur}  \\
    \midrule
    \(^2D_{3/2} - ^2P_{1/2}\) 	&	256~(35) & 162	\\
    \(^2P_{1/2} - ^2P_{3/2}\) 	&	154~(30) &	105 \\
    \(^2P_{3/2} - ^2P_{3/2}\) \text{ and } \(^2P_{1/2} - ^2P_{1/2}\)	&	192~(3)	&  \\
    \(^2P_{3/2} - ^2P_{1/2}\)	&	140~(17) & 83	\\
    \(^2S_{1/2} - ^2P_{3/2}\)	&	208~(28) &	196 \\
    \midrule
     \multicolumn{2}{c}{Argon} 	\\\midrule
    \(^2D_{5/2} -^2P_{3/2}\)	&	248(42)	 & 185 \\
    \(^2P_{1/2} -^2P_{1/2}\) and \(^2D_{3/2} - ^2P_{1/2}\)	&	617~(40) &	\\
    \(^2P_{1/2} -^2P_{1/2}\)	&	221~(8)	& 141 \\
    \(^2P_{3/2} -^2P_{3/2}\)	&	182~(8)	& 120 \\
    \(^2P_{3/2} -^2P_{1/2}\)	&	277~(72) & 120	\\
    \bottomrule
    \bottomrule
    \end{tabular}
    \end{table}

\subsection{Evaluation of experimental transition energies}\label{subsec:energy_determ}

Once the width of each contribution is determined, we can deduce the energy of every transition. The Bragg angle depends on the temperature of the crystals. However, the poor stabilization of the crystals temperature caused by the hot summer and the absence of a room temperature stabilization system, made the crystals drift in temperature throughout the day. This deviation is about \qty{1}{\celsius}, value which we use for the uncertainty in Table~\ref{tab:systematics}. The systematical uncertainties are obtained by the same methods as described in Refs.~\cite{asgl2012,assg2014,mssa2018}. 

\begin{table*}
    \centering
    \caption{List of systematic uncertainties for each spectrum. The systematics are separated into two categories. The first one, the daily systematic uncertainties, regroup all the systematics which can change on a daily basis. The overall contributions are systematics which are common for all spectra, such as the crystal parameters, uncertainties tied to the crystals rocking curves or the geometry of the setup.}\label{tab:systematics}
    									
    \begin{tabular}{lll}	
    \toprule									
    	Daily Contributions (eV)		&		Value (S)		&	Value (Ar) \\\midrule
    	Angular encoder error (0.2'')		&		0.0016		&		0.0035	\\
    	2nd crystal temperature stabilization (1.0\,°C)		&		0.0062		&		0.0079	\\
    	Variation of x-ray source size from 6 mm to 12 mm 		&		0.00462		&		0.0013	\\\midrule
    	Total	(eV)		&	0.0124	&		0.0127	\\
	\toprule
    	Overall Contributions (eV)		&	\text{	Value (S)	}	&	\text{Value (Ar)} \\\midrule	
    	Temperature reading (0.5\,°C)		&		0.0031		&		0.0040	\\		
    	Vertical tilts of crystals (±0.01°) for each crystal		&		0.00085		&		0.0002	\\
    	Vertical divergence (1 mm)		&		0.00102		&		0.0002	\\
    	Si crystal atomic form factor		&		0.003		&		0.002	\\
    	X-ray polarization		&		0.006		&		0.0014	\\
    	Lattice spacing error		&		0.0000287		&		0.000037	\\
    	Index of refraction		&		0.00055		&		0.0016	\\
    	Thermal expansion		&		0.00015		& 0.00019\\\midrule
    	Total	(eV)	&		0.0075		&		0.0069	\\\bottomrule
    \end{tabular}									
    \end{table*}

\subsubsection{Temperature correction}

\begin{figure*}
    \caption{Effect of the temperature correction on the energy for \(1s 2s^2 2p^2~^{2}P_{3/2} - 1s2~2s^2~2p~^2P_{3/2}\) argon (left) and \(1s 2s^2 2p^2~^{2}P_{1/2} - 1s2~2s^2~2p~^2P_{1/2}\) and \(1s 2s^2 2p^2~^{2}P_{3/2} - 1s2~2s^2~2p~^2P_{3/2}\) sulfur (right). In sulfur, this correction reduces the day-per-day scattering. In argon, this correction reduces the scattering of the measurements compared to the measured value. The energy value is here calculated with a weighted average method, with statistical error (light red) and with the systematical uncertainties in dashed orange lines.}\label{fig:energy_correction}
    \includegraphics[width=.5\linewidth]{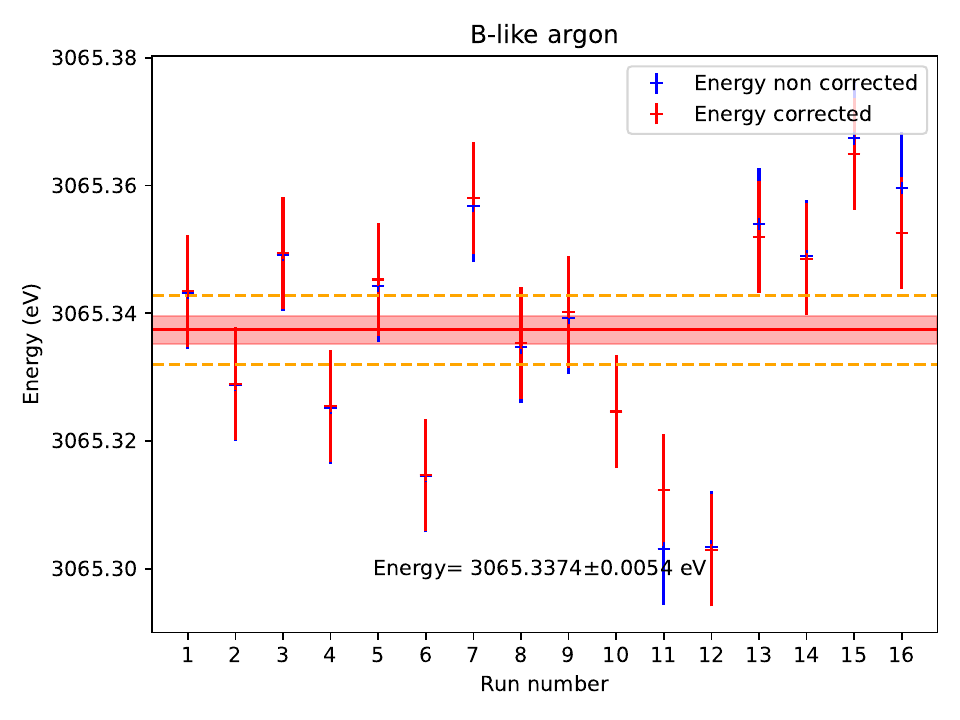}
    \includegraphics[width=.5\linewidth]{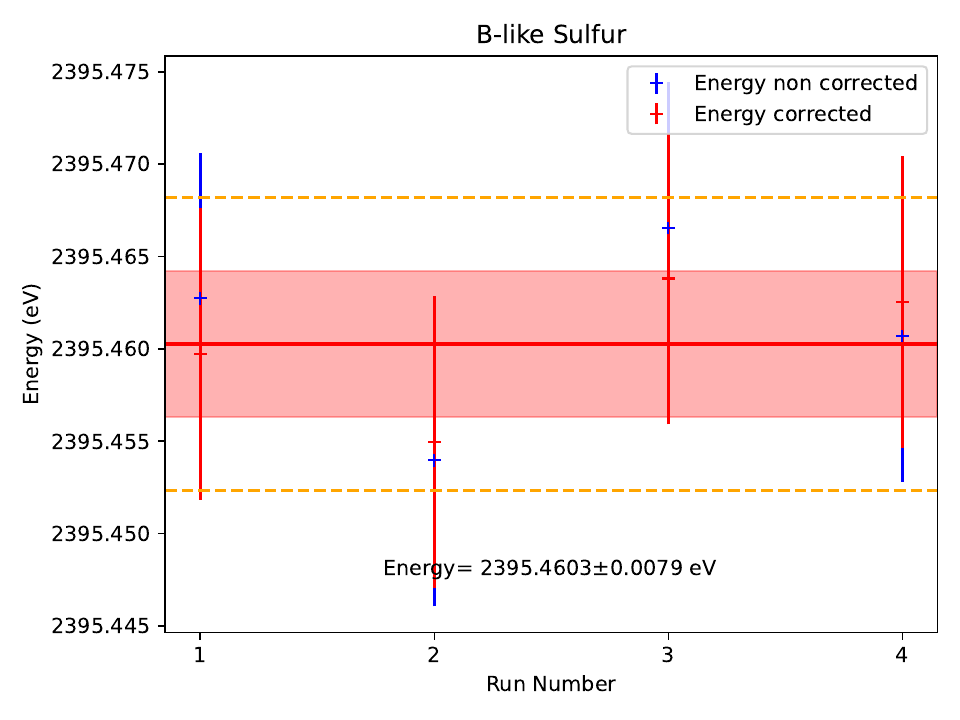}
\end{figure*}

With the natural linewidth determined above, new sets of simulations of both dispersive and nondispersive spectra are performed using the corresponding linewidth, for different transitions energy and crystal temperature, which shifts the line position. We use five different energies \(E_k=E+k\Delta E\), with \(\Delta E\approx\qty{10}{\milli\electronvolt}\), \(E\) being the energy deduced from the difference in angle from the width analysis, and ten temperatures from 22.5\,°C to 27\,°C for each line. The simulations were performed using the same temperature for both crystals. We fit again by applying the peak per peak cut and the neighbor modelling
However, the temperature of the crystals are different and changes during the day. The only shift not taken into account by our simulations is the temperature difference between the two crystals. The angle difference is caused by the dilatation of the crystals, caused by the temperature change, changing the lattice spacing of the crystals. This dilatation is as following \cite{sab2001}:
\begin{align}
    \frac{\Delta d}{d}= \eta_0(T-20) + \eta_1(T-20)^2
\end{align}
with T the crystal temperature, $\eta_0$ and $\eta_1$ the coefficient of dilatation of the crystal. Using this the Bragg angle can be written as:
\begin{align}\label{eq:Bragg_law}
    \Theta_{\text{Bragg}}(E,T)= \arcsin\left(\frac{n\lambda(1-\delta(E))}{2d(T)}\right)
\end{align}
with d(T) the lattice spacing of the crystal at temperature T, \(\delta(E)\) the refractive index correction and \(\Theta_{\text{Bragg}}\) the Bragg angle of a photon of energy \(E\).
The angle difference induced by the change of the lattice spacing of the first crystal, due to the difference of temperature between both crystals  can be well approximated by using Eq.~\eqref{eq:Bragg_law}:
\begin{align}\label{eq:angle_corr}
    \Delta\theta&=\Theta_{\text{Bragg}}(E,T_{\text{cryst 1}})-\Theta_{\text{Bragg}}(E,T_{\text{cryst 2}})
\end{align}
This correction is applied for each spectrum, both dispersive and nondispersive, and reduce the overall scattering of our data along the measurements as shown in Fig.~\ref{fig:energy_correction}.  

The line energy is obtained from the two bi-dimensional fits from the relation \(\Delta\theta_{\text{Expt-simul}}(E_{\text{Expt}},T_{\text{Expt}})=0\) where \(T_\text{Expt}\) is the average temperature of the second crystal, to take into account the temperature of the second crystal.

\subsubsection{Determination of the energy}
\begin{figure*}
    \caption{Determination of the energy using log-likelihood for sulfur. Each measurement point is including the daily uncertainties, and the final energy uncertainty, the overall uncertainty is added. Among all measurement, weighted average and Bayesian average are compatible, and the uncertainty of the Bayesian average is higher than the weighted mean. }\label{fig:energy_loglikelihood_s}
\begin{subfigure}{.5\textwidth}
    \caption{\(1s 2s^2 2p^2~^2D_{3/2}-1s^2 2s^2 2p~^2P_{1/2}\)}\label{fig:s_p1}
    \includegraphics[width=\linewidth]{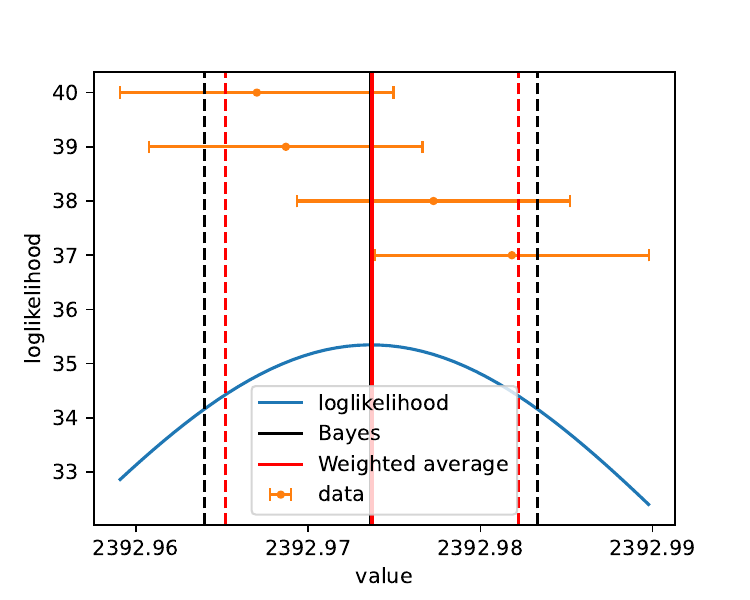}
\end{subfigure}
\begin{subfigure}{.5\textwidth}
    \caption{\(1s 2s^2 2p^2~^2P_{1/2}-1s^2 2s^2 2p~^2P_{3/2}\) }\label{fig:s_p2}
    \includegraphics[width=\linewidth]{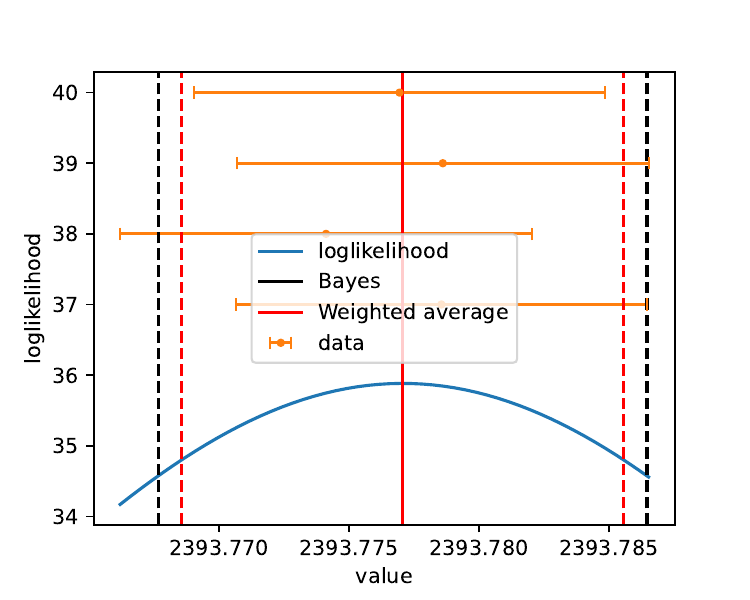}
\end{subfigure}

\begin{subfigure}{.5\textwidth}
    \caption{\(1s 2s^2 2p^2~^2P_{1/2}-1s^2 2s^2 2p~^2P_{1/2}\) and \(1s 2s^2 2p^2~^2P_{3/2}-1s^2 2s^2 2p~^2P_{3/2}\)}\label{fig:s_p3}
    \includegraphics[width=\linewidth]{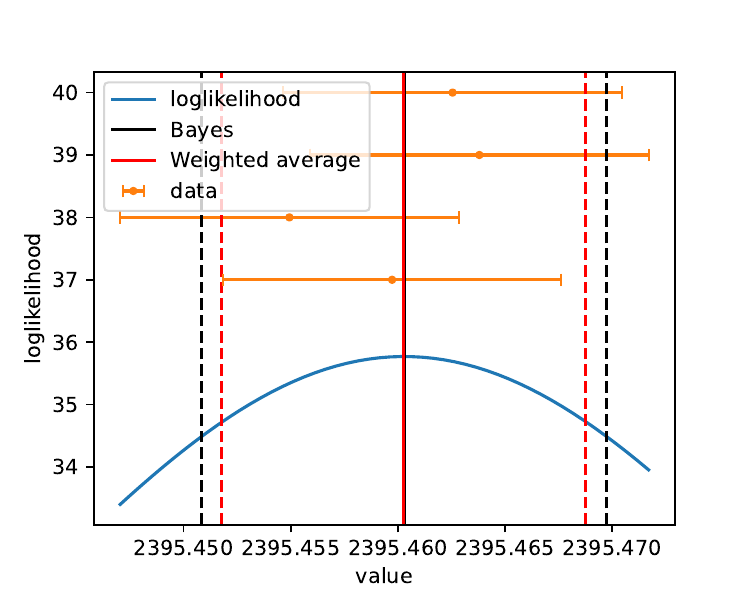}
\end{subfigure}
\begin{subfigure}{.5\textwidth}
    \caption{\(1s 2s^2 2p^2~^2P_{3/2}-1s^2 2s^2 2p~^2P_{1/2}\)}\label{fig:s_p4}
    \includegraphics[width=\linewidth]{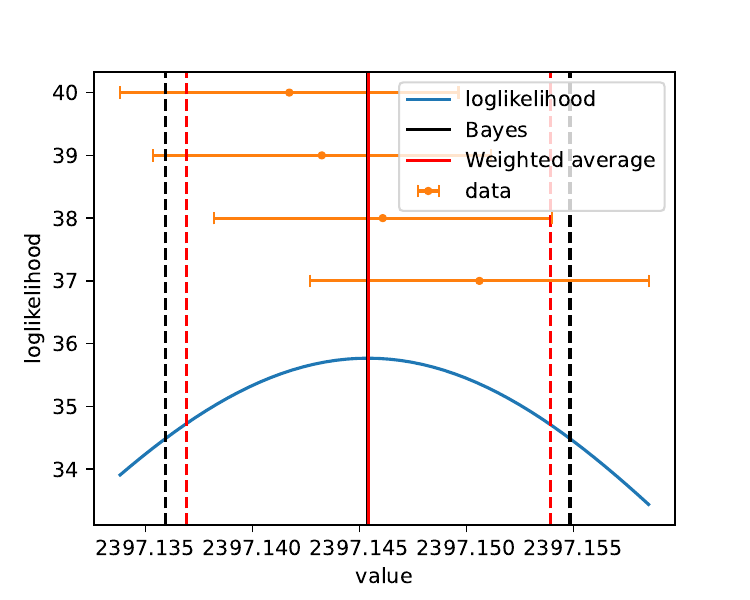}
\end{subfigure}

\begin{subfigure}{.5\textwidth}
    \caption{\(1s 2s^2 2p^2~^2S_{1/2}-1s^2 2s^2 2p~^2P_{3/2}\)}\label{fig:s_p5}
    \includegraphics[width=\linewidth]{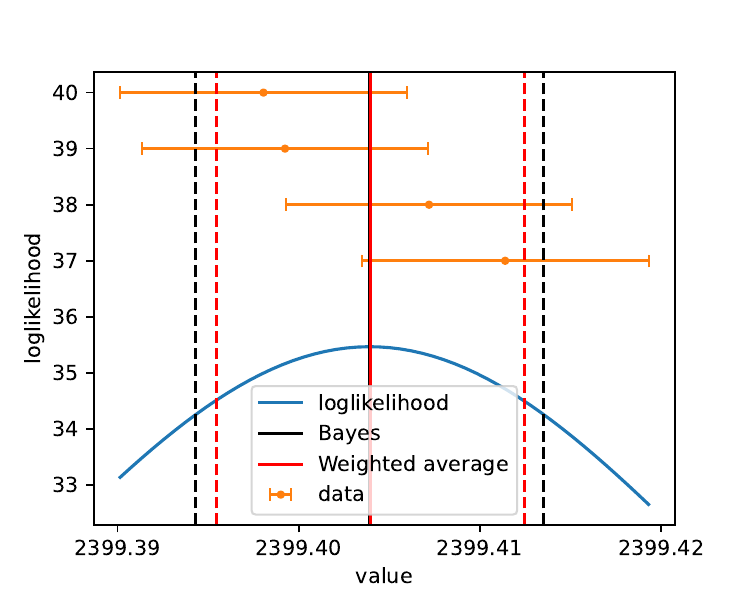}
\end{subfigure}
\end{figure*}

\begin{figure*}
    \caption{Determination of the energy using log-likelihood for argon. Each measurement point is including the daily uncertainties, and the final energy uncertainty, the overall uncertainty is added. Among all measurement, weighted average and Bayesian average are compatible. Nonetheless, the outliers measurements, due to low statistics are shifting the total measured energy, whereas the Bayesian are excluding them. The Bayesian average approach also allows to adapt the uncertainty on the scattering of the points, whereas the statistical uncertainty from the weighted average does not depend on the points themselves, and rely on thumb rules in case of large scattering.}\label{fig:energy_loglikelihood_ar}
\begin{subfigure}{.5\textwidth}
    \includegraphics[width=\linewidth]{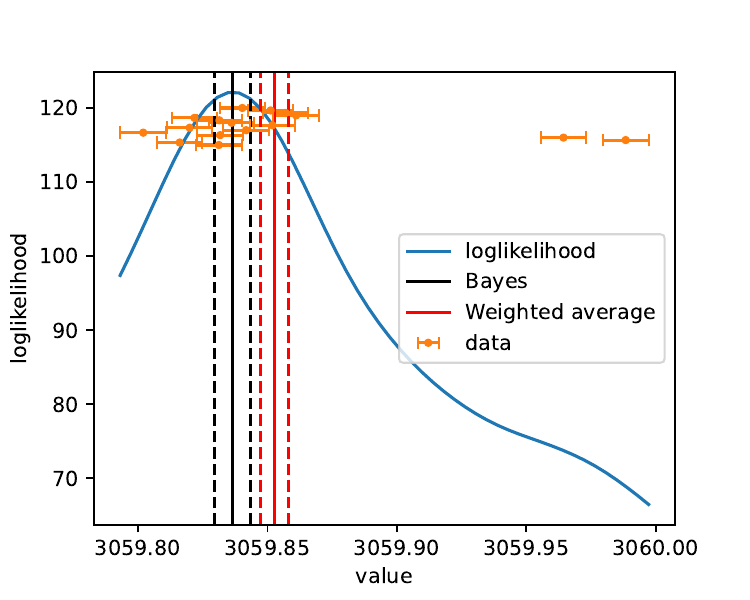}
    \caption{\(1s~2s^2~2p^2~^2D_{5/2}-1s^2~2s^2~2p~^2P_{3/2}\)}\label{fig:ar_p1}
\end{subfigure}
\begin{subfigure}{.5\textwidth}
    \includegraphics[width=\linewidth]{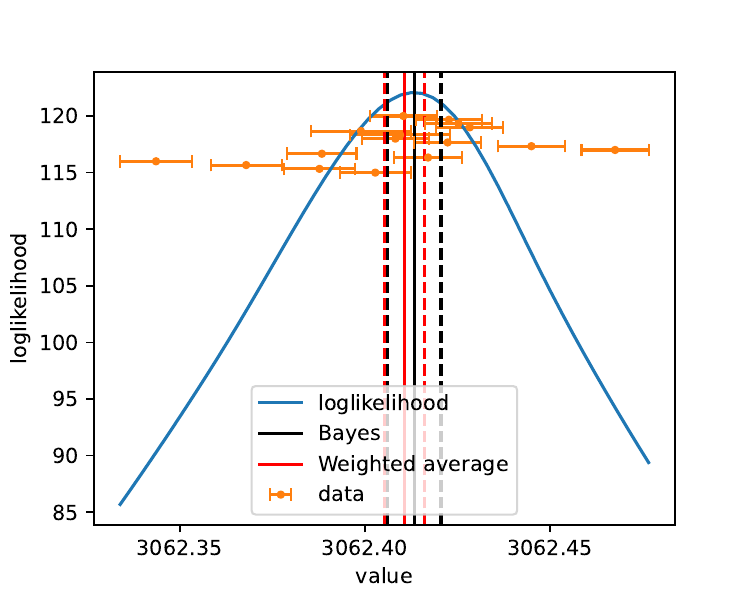}
    \caption{\(1s~2s^2~2p^2~^2D_{3/2}-1s^2~2s^2~2p~^2P_{1/2}\) and \(1s~2s^2~2p^2~^2P_{1/2}-1s^2~2s^2~2p~^2P_{3/2}\)}\label{fig:ar_p2}
\end{subfigure}
\begin{subfigure}{.5\textwidth}
    \includegraphics[width=\linewidth]{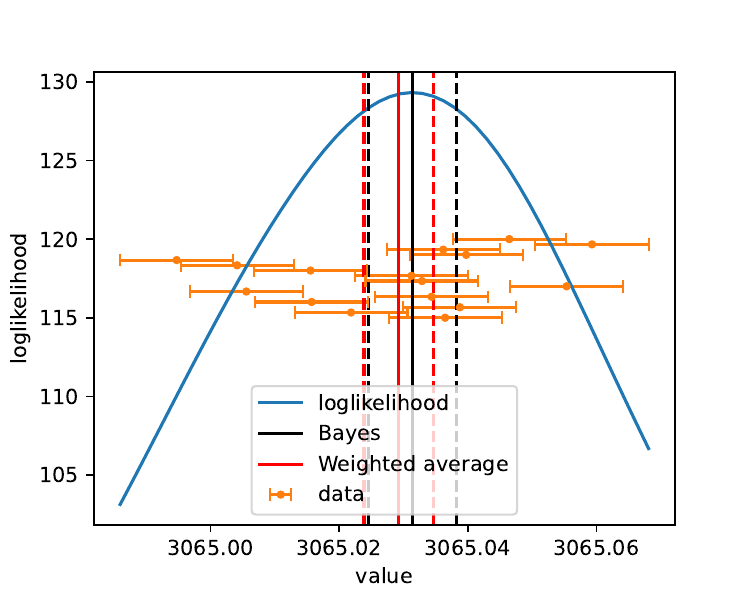}
    \caption{\(1s~2s^2~2p^2~^2P_{1/2}-1s^2~2s^2~2p~^2P_{1/2}\)}\label{fig:ar_p32}
\end{subfigure}
\begin{subfigure}{.5\textwidth}
    \includegraphics[width=\linewidth]{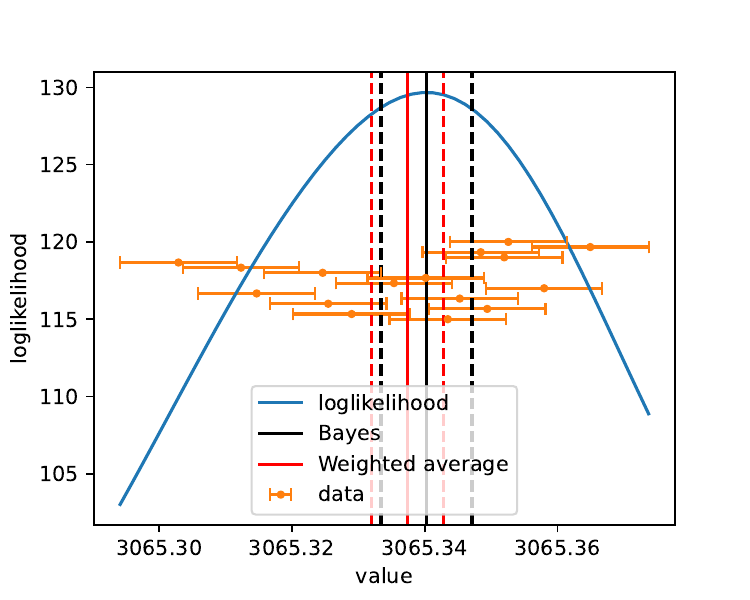}
    \caption{\(1s~2s^2~2p^2~^2P_{3/2}-1s^2~2s^2~2p~^2P_{3/2}\)}\label{fig:ar_p3}
\end{subfigure}
\begin{subfigure}{.5\textwidth}
    \includegraphics[width=\linewidth]{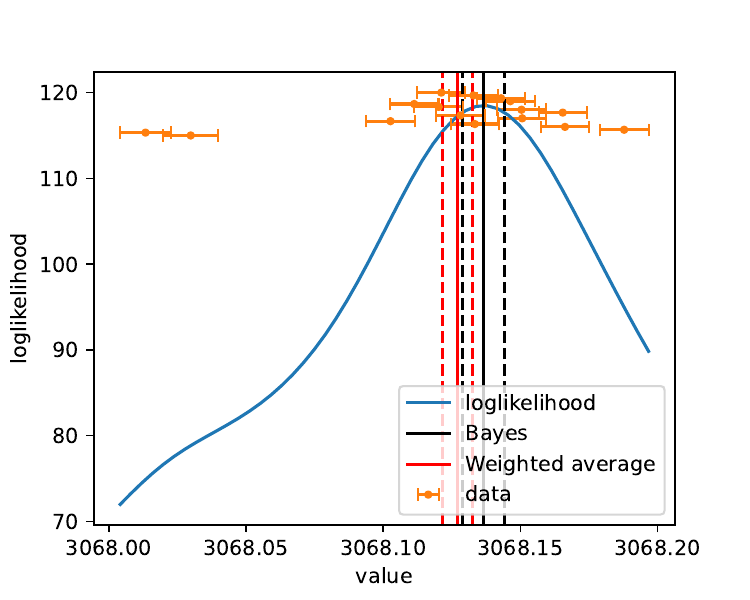}
    \caption{\(1s~2s^2~2p^2~^2P_{3/2}-1s^2~2s^2~2p~^2P_{1/2}\)}\label{fig:ar_p4}
\end{subfigure}
\begin{subfigure}{.5\textwidth}
    \includegraphics[width=\linewidth]{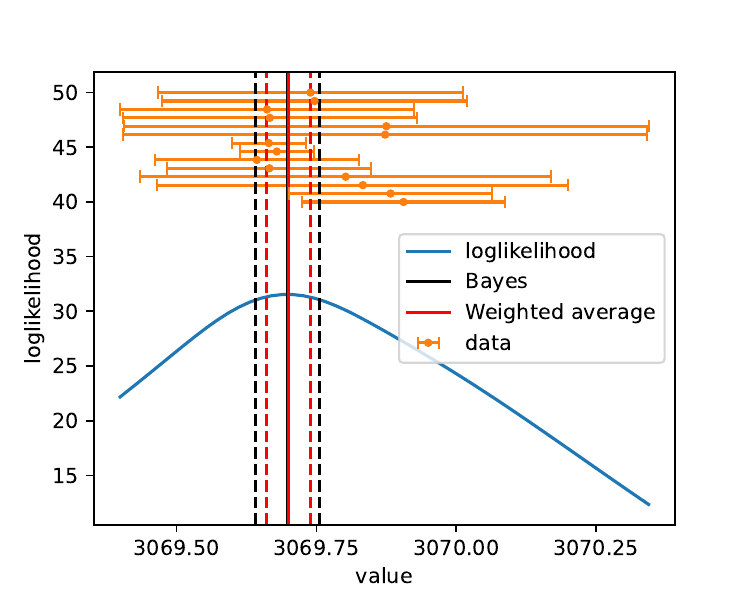}
    \caption{\(1s~2s^2~2p^2~^2S_{1/2}-1s^2~2s^2~2p~^2P_{3/2}\)}\label{fig:ar_p5}
\end{subfigure}
\end{figure*}

The energies deduced from each run can now be processed, in order to obtain the final energy value and its uncertainty. Our data are much more scattered than previous measurements \cite{assg2014,mssa2018,mbpt2020,mpsd2023}. The usual way to process the data is by doing a inverse-variance weighted average. This method relies on the correct determination of the uncertainties of the values to weight~\cite{zbbd2020}. However, when inconsistent data have to be considered (with a scattering much larger than the typical size of the assigned error bars), the final uncertainty of the weighted average does not depend on the scattering of the points with the risk of its underestimation. In our case, compared to the previous measurement from the DCS, our daily measurements suffer from a large scattering, making the use of the standard weighted average unappropriated.
In order to resolve this problem, we perform a Bayesian average (BAv)~\cite{tam2024,sas2006}, where the uncertainty related to each data point is considered as a lower bound of the real uncertainty including unknown systematic errors. For each measurement, day-dependent systematical errors are added to each measurement point, before performing the log-likelihood procedure. To get the final uncertainty, the systematical uncertainties do not depend on the daily measurement conditions are added to the log-likelihood uncertainty. In this analysis, and especially for the argon spectra, where the measurements have outliers and high scattering, the BAv allows to estimate correctly the statistical uncertainty to the scattering of the data and to be robust with respect to the possible presence of outliers, which shifts the measured energy. Outliers were observed in argon spectra, especially for the \(1s 2s^2 2p^2~^{2}D_{5/2} - 1s2~2s^2~2p~^2P_{3/2}\)  and \(1s 2s^2 2p^2~^{2}S_{1/2} - 1s2~2s^2~2p~^2P_{3/2}\) argon (cf Fig.~\ref{fig:energy_loglikelihood_ar}). Since only 2 dispersive spectra have been taken for sulfur, we do not observe outliers (cf Fig.~\ref{fig:energy_loglikelihood_s}).

\begin{table}[hb]
    \centering
    \caption{Experimental energy of the transitions in Ar and S. The \(1s 2s^2 2p^2~^2P_{1/2} -1s 2s^2 2p^2~^2P_{1/2}\) and  \(1s 2s^2 2p^2~^2D_{3/2} - 1s^2 2s^2 2p~^2P_{1/2}\) transitions in Ar couldn't be resolved.}\label{tab:energy_full}
    \begin{tabular}{ld}\toprule
    Transition	&	\multicolumn{1}{c}{Energy ~(eV)}	\\
    \midrule
     \multicolumn{2}{c}{Sulfur} \\
    \midrule
     \(1s 2s^2 2p^2~^2D_{3/2} - 1s^2 2s^2 2p~^2P_{1/2}\) &	2392.9736~(92)  \\
     \(1s 2s^2 2p^2~^2P_{1/2} - 1s^2 2s^2 2p~^2P_{3/2}\)*	&	2393.7771~(89)  \\
     \(1s 2s^2 2p^2~^2P_{1/2} - 1s^2 2s^2 2p~^2P_{1/2}\)* &   2395.4603~(89) \\ 
     \(1s 2s^2 2p^2~^2P_{3/2} - 1s^2 2s^2 2p~^2P_{3/2}\) &	2395.4603~(89) \\
     \(1s 2s^2 2p^2~^2P_{3/2} -1s 2s^2 2p^2~^2P_{1/2}\)	&	2397.1454~(89)  \\
     \(1s 2s^2 2p^2~^2S_{1/2} -1s 2s^2 2p^2~^2P_{3/2}\)	&	2399.4039~(91)  \\  
       \midrule
     \multicolumn{2}{c}{Argon} \\
    \midrule
   \(1s 2s^2 2p^2~^2D_{5/2} - 1s^2 2s^2 2p~^2P_{3/2}\)	&	3059.8366~(70) \\
    \(1s 2s^2 2p^2~^2P_{1/2} - 1s^ 2s^2 2p~^2P_{1/2}\)*   &   3062.4133~(73) \\ 
    \(1s 2s^2 2p^2~^2D_{3/2} - 1s^2 2s^2 2p~^2P_{1/2}\)*	&	3062.4133~(73) \\
    \(1s 2s^2 2p^2~^2P_{1/2} - 1s^2 2s^2 2p~^2P_{1/2}\)	&	3065.0314~(69)  	\\
    \(1s 2s^2 2p^2~^2P_{3/2} - 1s^2 2s^2 2p~^2P_{3/2}\)	&	3065.3403~(69)  	\\
    \(1s 2s^2 2p^2~^2P_{3/2} - 1s^2 2s^2 2p~^2P_{1/2}\)	&	3068.1366~(77)  	\\
    \(1s 2s^2 2p^2~^2S_{1/2} - 1s^2 2s^2 2p~^2P_{3/2}\)  &   3069.698~(57) \\\botrule
    \bottomrule
    \end{tabular}
\flushleft
 * Unresolved.
    \end{table}

\section{Theoretical calculations}
\label{sec:theo-calc}
In order to identify the transition observed, and to test relativistic many-body techniques and QED contributions for a core-excited 5-electron systems, we have performed Multi-Configuration Dirac-Fock calculations of all possible  $1s 2s^2 2p^2 \, ^{2S+1}L_J$ levels and of the two $1s^2 2s^2 2p \, ^2P_J$ levels for argon and sulfur. We also performed calculations for these levels on the few other ions for which measurements are available, C$^+$, Si$^{9+}$, Cr$^{19+}$ and Fe$^{21+}$.
The calculations have been performed using the 2024 version of the \textsc{mdfgme} code \cite{des1975,des1993,iad2005,ind2024}. The calculation was performed using single and double excitations up to principal quantum number $n=4$ or $n=5$ depending on convergence issues, in the Optimized Level (OL) scheme, \ie all orbitals are fully relaxed. For the two lower levels, triple excitations were also included to compare with accurate experiments, in particular the laser measurement from Refs \cite{mkbl2011,mlkb2020}. Accurate values for the fine structure interval of the the two $1s^2 2s^2 2p \, ^2P_J$ levels were helpful in the identification of the observed lines, as several lines decays to both levels and should have energy differences to the ground state fine structure splitting.

In these calculations, the Breit interaction is included to all orders \cite{iad1993}. Finite nuclear size is included, with a Fermi model, using mean spherical radii from \cite{aam2013}. The self-energy with finite nuclear size is included for $s$ and $p_{1/2}$ levels and for point nuclei for all other levels \cite{moh1974,moh1992,mak1992,iam1998,iam1998a}. The self-energy screening is included using the model operator from \cite{sty2013,sty2015}. The Uelhing vacuum polarization is included to all orders \cite{ind2013}. The Wichmann and Kroll and higher-order corrections as well as two-loop corrections are included \cite{yis2003,yis2003a,yis2005,yis2005a,yis2008}. Projection operators are taken into account for all correlation orbitals\cite{ind1995}, and all the single excitations, including those relevant from the Brillouin's theorem are taken into account to insure a correct non-relativistic limit \cite{ild2005}.

The radiative and Auger rate are evaluated with the \textsc{mdfgme}, including only the largest contributions, due to intrashell correlation. In both cases the initial and final state wavefunctions are fully relaxed and the non-orthogonality between those wavefunctions are fully accounted for as described in Refs. \cite{low1955,hag1978,ind1996,ind1997}. The width is largely dominated by a few Auger transitions from the initial state to the Be-like $1s^2 2p^2\, ^3P_0$ and $1s^2 2s 2p \,^3P_J$ states, with $J=0, 1$ and $2$. Auger transitions to more excited states like $1s^2 2s n\ell$, with $n\geq 3$ represents less than  \qty{1}{\percent} of the width. The radiative transitions represent less than \qty{0.1}{\percent} of the width.

\begin{figure*}
\caption{Comparison theory experiment for all existing $2p \to 1s$ transitions in B-like ions with a $1s$ hole, scaled by $Z^2$. }
\label{fig:comp_th_exp_all_z}
    \includegraphics[width=.9\linewidth]{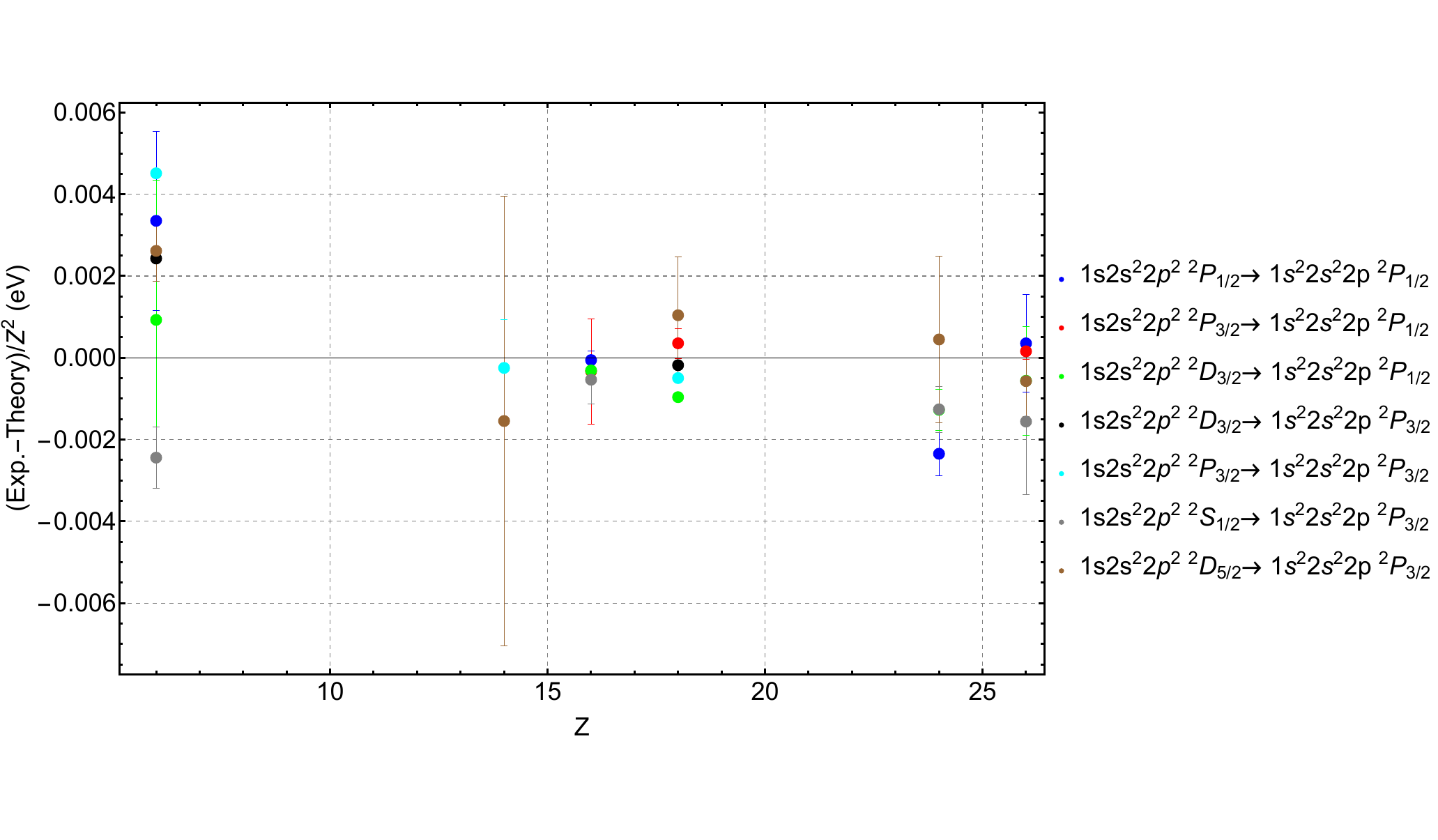}
\end{figure*}

The comparison between the present calculations and all the measurements available to date are shown on Fig. \ref{fig:comp_th_exp_all_z}.

\section{Discussion and comparison with theory}
\label{sec:disc-comp-theo}
According to NIST database, most of the transitions  identified in the present work have not been measured yet, with the exception of two transition in argon. For \(1s 2s^2 2p^2~^2P_{3/2} - 1s^2 2s^2 2p~^2P_{3/2}\) and \(1s 2s^2 2p^2~^2D_{3/2} - 1s^2 2s^2 2p~^2P_{1/2}\) in argon, our measurement is in agreement with the measurement of Magunov~\cite{mfsp2002} within their \qty{0.4}{\electronvolt} uncertainty. An EBIT measurement for B-like sulfur have been performed by Hell~\cite{hbwg2016}, but the level were not separated enough to provide identification. Comparison with theory is thus crucial to ensure right identification of the levels.  We compared our results with MultiConfigurational Dirac Foch (MCDF) calculations using \textsc{mdfgme}. The multiconfigurational approach allows taking into consideration a large amount of electronic correlation by taking into account a limited amount of electronic configurations. Here have been included all singly and doubly excited states up to 4f orbitals. The calculations with triply excited states  were unstable and level crossing were occurring, shifting away the energy of the level. The results of our calculations and their comparison to the experimental data are shown in Table~\ref{tab:ar_comp_th} for argon and Table~\ref{tab:s_comp_th} for sulfur. These results are particularly interesting to compare to theory as the coupling is changing from LS to JJ coupling in this region. We compared our theoretical results with other calculations using different methods such as relativistic Hartree-Fock method, by Biémont~\cite{bqfs2000}, Magunov~\cite{mfsp2002}, Palmeri~\cite{pqmb2008}, relativistic Dirac-Fock method (using \textsc{fac}~\cite{gu2008}) performed by Hell~\cite{hbwg2016}, Dirac-Fock from Costa \cite{cmps2001} and multiconfigurational Dirac-Fock from Biémont~\cite{bqfs2000}. The inclusion of the correlation is not done for the Relativistic Hartree-Fock (HFR) and \textsc{fac} calculations, which can explain the large discrepancies between the calculated energies and the measurement, as the correlations can shift the level energies by few \qty{0.1}{\electronvolt}. 

Our measurement is overall consistent with the MCDF calculations. However, several high precision measurements of the \(1s^2 2s^2 2p~^2P_{3/2} - ^2P_{1/2}\) from laser spectroscopy of B-like argon and grating spectroscopy of the sun~\cite{jeff1969} differs from our observation of the \(1s^2 2s^2 2p~^2P_{3/2} - ^2P_{1/2}\). 

\begin{table*}
\centering
    \caption{Comparison between the experimental and the theoretical energy of the sulfur transitions. This difference is compared to the addition of both theoretical and experimental uncertainties. All experimental and theoretical data are in good agreement.}\label{tab:s_comp_th}
        \begin{tabular}{cccccccc}
            \toprule
            Initial Level			&	Final level			& Exp. energy	 &	MCDF (This Work)		&	\(E_{\textrm{Exp.}}-E_{\textrm{Theo.}}\)	\\\midrule
    \(1s 2s^2 2p^2~^2D_{3/2}\) 	& \(1s^2 2s^2 2p^1~^2P_{1/2}\) & 2392.9736(92) & 2393.0527(20)  & 0.079(22) \\
    \(1s 2s^2 2p^2~^2P_{1/2}\) 	& \(1s^2 2s^2 2p^1~^2P_{3/2}\) & 2393.7770(89) & 2393.825(48)   & 0.048(49) \\
    \(1s 2s^2 2p^2~^2P_{1/2}\) 	& \(1s^2 2s^2 2p^1~^2P_{1/2}\) & 2395.4603(89)* & 2395.475(20)   & 0.015(22) \\
    \(1s 2s^2 2p^2~^2P_{3/2}\) 	& \(1s^2 2s^2 2p^1~^2P_{3/2}\) & 2395.4603(89)* & 2395.561(55)   & 0.100(55) \\
    \(1s 2s^2 2p^2~^2P_{3/2}\) 	& \(1s^2 2s^2 2p^1~^2P_{1/2}\) & 2397.1454(89) & 2397.210(56)   & 0.065(56) \\
    \(1s 2s^2 2p^2~^2S_{1/2}\) 	& \(1s^2 2s^2 2p^1~^2P_{3/2}\) & 2399.4039(91) & 2399.541(15)   & 0.137(18) \\
    \bottomrule
        \end{tabular}
        \flushleft * Unresolved.
    \caption{Comparison between the experimental and the theoretical energy of the argon transitions. This difference is compared to the addition of both theoretical and experimental uncertainties. All experimental and theoretical data are in good agreement, except for the blended \(1s 2s^2 2p^2~^2D_{3/2}-1s^2 2s^2 2p~^2P_{1/2}\).}\label{tab:ar_comp_th}
    \centering
    \begin{tabular}{ccccc}
            \toprule
            Initial Level			&	Final level		&	Exp. energy	&	MCDF (This Work)	&	\(E_{\textrm{Exp.}}-E_{\textrm{Theo.}}\) \\\midrule
        \(1s 2s^2 2p^2~^2D_{5/2}\) 	& \(1s^2 2s^2 2p~^2P_{3/2}\) & 3059.8366(70) & 3059.50(45)  &  -0.34(45) \\
        \(1s 2s^2 2p^2~^2P_{1/2}\) 	& \(1s^2 2s^2 2p~^2P_{3/2}\) & 3062.3014(69)\footnotemark[1] & 3062.260(61) &  -0.041(62)\\
        \(1s 2s^2 2p^2~^2D_{3/2}\) 	& \(1s^2 2s^2 2p~^2P_{1/2}\) & 3062.4133(73) & 3062.725(24) &  0.312(25) \\
        \(1s 2s^2 2p^2~^2P_{1/2}\) 	& \(1s^2 2s^2 2p~^2P_{1/2}\) & 3065.0314(69) & 3065.090(62) &  0.058(62) \\
        \(1s 2s^2 2p^2~^2P_{3/2}\) 	& \(1s^2 2s^2 2p~^2P_{3/2}\) & 3065.3403(69) & 3065.19(11)  &  -0.15(11) \\
        \(1s 2s^2 2p^2~^2P_{3/2}\) 	& \(1s^2 2s^2 2p~^2P_{1/2}\) & 3068.1366(77) & 3068.02(11)  &  -0.12(11) \\
        \(1s 2s^2 2p^2~^2S_{1/2}\) 	& \(1s^2 2s^2 2p~^2P_{3/2}\) & 3069.698(57) & 3069.833(75)  &  0.135(57)  \\\bottomrule
        \end{tabular}
        \flushleft
        * Unresolved \\
        $^1$ Value calculated from the energy of \(1s 2s^2 2p^2~^2P_{1/2} - 1s^2 2s^2 2p~^2P_{1/2}\) minus the energy of \(1s^2 2s^2 2p^2~^2P_{1/2} - 1s^2 2s^2 2p~^2P_{1/2}\) from NIST database~\cite{krrn2023}.
        \end{table*}

We also performed the calculations of the \(1s 2s^2 2p^2~^2P_{1/2} - 1s^2 2s^2 2p~^2P_{1/2}\), \(1s 2s^2 2p^2~^2P_{3/2} - 1s^2 2s^2 2p~^2P_{3/2}\) and \(1s 2s^2 2p^2~^2S_{1/2} - 1s^2 2s^2 2p~^2P_{3/2}\)  for other B-like systems that were measured, such as carbon~\cite{kah2022}, chromium~\cite{bfq1999}, and iron~\cite{bpjh1993,rbes2013}. The calculations are overall in good agreement with the measurements.

\section{Conclusion}
We reported the measurement of $2p \to 1s$ transitions from argon and sulfur. We were able to resolve line blending using Bayesian analysis, and we report a precision of 4 ppm in sulfur and 2 ppm in argon for the measurement of the $1s 2s^2 2p^2~^2P_i - 1s^2 2s^2 2p~^ 2P_i,~i \in {1/2,3/2}$ transitions. We also performed high-precision calculations for transitions in K-hole transition from carbon up to iron with good agreement with experimental values.

\begin{table*}\label{tab:s_comp_th_th}
    \caption{Comparison between different calculation, from Hell~\cite{hbwg2016} (FAC 16), Palmeri (HFR 08) ~\cite{pqmb2008} and this work, using MCDF. Differences between the calculations are up to \qty{1}{\electronvolt}. }
    \centering
        \begin{tabular}{cccccc}
            \toprule
            Upper level & Lower level & FAC 16 & HFR 08 & MCDF (this work) & \\\midrule
\(1s 2s^2 2p^2~^2D_{5/2}\) & \(1s^2 2s^2 2p~^2P_{3/2}\) & 2391.41 & 2391.6871 & 2391.2998 \\
\(1s 2s^2 2p^2~^2D_{3/2}\) & \(1s^2 2s^2 2p~^2P_{1/2}\) & 2393.07 & 2393.0754 & 2393.0527 \\
\(1s 2s^2 2p^2~^2P_{1/2}\) & \(1s^2 2s^2 2p~^2P_{3/2}\) & 2393.50 & 2394.0041 & 2393.8254 \\
\(1s 2s^2 2p^2~^2P_{1/2}\) & \(1s^2 2s^2 2p~^2P_{1/2}\) & 2395.11 & 2395.6241 & 2395.4748 \\
\(1s 2s^2 2p^2~^2P_{3/2}\) & \(1s^2 2s^2 2p~^2P_{3/2}\) & 2395.25 & 2395.6854 & 2395.5610 \\
\(1s 2s^2 2p^2~^2P_{3/2}\) & \(1s^2 2s^2 2p~^2P_{1/2}\) & 2396.87 & 2397.3054 & 2397.2104 \\
\(1s 2s^2 2p^2~^2S_{1/2}\) & \(1s^2 2s^2 2p~^2P_{3/2}\) &  \dots  & 2399.8542 & 2399.5411 \\\bottomrule
        \end{tabular}
    \end{table*}

\begin{table*}\label{tab:ar_comp_th_th}
\caption{Comparison between different calculation, using Hartree-Fock relativistic method from Palmeri~\cite{pqmb2008} (HFR 08), Biémont~\cite{bqfs2000} (HFR 00), Dirac-Fock method from Costa~\cite{cmps2001} (DF 01) and MultiConfigurational Dirac-Fock (MCDF) from this work. The use of level correlation shifts the atomic lies by more than few \qty{100}{\milli\electronvolt}. Calculation of the correlations were critical for the identification of the transitions.}
\centering
    \begin{tabular}{cccccc}
        \toprule
        Upper level & Lower level & HFR08 & HFR 00 & DF 01 & MCDF (This work)\\\midrule
        \(1s~2s^2~2p^2~^2D_{5/2}\) & \(1s^2 2s^2 2p~^2P_{3/2}\) & 3060.623 & 3059.978 & 3061.036 & 3059.499\\
        \(1s~2s^2~2p^2~^2D_{3/2}\) & \(1s^2 2s^2 2p~^2P_{1/2}\) & 3062.893 & 3062.246 & 3063.683 & 3062.725\\
        \(1s~2s^2~2p^2~^2P_{1/2}\) & \(1s^2 2s^2 2p~^2P_{3/2}\) & 3062.724 & 3061.867 & 3062.851 & 3062.260\\
        \(1s~2s^2~2p^2~^2P_{1/2}\) & \(1s^2 2s^2 2p~^2P_{1/2}\) & 3065.523 & 3064.744 & 3065.653 & 3065.090\\
        \(1s~2s^2~2p^2~^2P_{3/2}\) & \(1s^2 2s^2 2p~^2P_{3/2}\) & 3065.752 & 3064.971 & 3066.108 & 3065.192\\
        \(1s~2s^2~2p^2~^2P_{3/2}\) & \(1s^2 2s^2 2p~^2P_{1/2}\) & 3068.550 & 3067.777 & 3068.840 & 3068.022\\
        \(1s~2s^2~2p^2~^2S_{1/2}\) & \(1s^2 2s^2 2p~^2P_{3/2}\) & 3070.311 & 3068.612 & 3069.980 & 3069.833\\\bottomrule
    \end{tabular}
\end{table*}

\subsection*{Acknowledgement}
L.D. thanks Daniel Pinheiro for useful discussions about the temperature correction. 

\subsection*{Author Contribution Statement}
The experiment was prepared and performed by P.I., E.L., S.M., J.M., C.P. and M.T. The simulations were performed by L.D. with the help of J.M. The data were analyzed by L.D. with the help of P.I., J.M., N.P. and M.T. The Bayesian average was developped by M.M. and M.T. Theoretical calculations were performed by P.I. All authors discussed and approved the data as well as the paper.

\textbf{Funding information}  P. Indelicato is a member of the Allianz Program of the Helmholtz Association, contract no EMMI HA-216 ‘‘Extremes of Density and Temperature: Cosmic Matter in the Laboratory''. He acknowledge support from CNRS INP for funding of the cluster used to run part of the presented calculations. We acknowledge support from the PESSOA Huber Curien Program 2022, No. 47863UE, and PAUILF Program No. 2017-C08. This research was funded in part by Fundação para a Ciência e Tecnologia (FCT; Portugal) through research center Grant No. UID/FIS/04559/2020 to LIBPhys-UNL from the FCT/MCTES/PIDDAC, Portugal. The SIMPA ECRIS has been financed by grants from CNRS, MESR, and University Pierre and Marie Curie (now Sorbonne Université). The DCS was constructed using grants from BNM 01 3 0002 and the ANR (Grant No. ANR-06-BLAN-0223). N.P. thanks the CNRS for support. L.D. thanks the Sorbonne University Institute Physics of Infinities and the research federation PLAS@PAR for his Ph.D. grant.

\textbf{Data Availability Statement}
This manuscript has no data or the data will not be deposited. Data will be made available from the corresponding author upon reasonable request.
\bibliography{refs}

\end{document}